\documentstyle[multicol,prl,aps,amsfonts]{revtex}

\newcommand{\Ga}{\alpha}
\newcommand{\Gb}{\beta}

\newcommand{\Gd}{\delta}

\newcommand{\Gg}{\gamma}
\newcommand{\GG}{\Gamma}

\newcommand{\Gl}{\lambda}
\newcommand{\GL}{\Lambda}
\newcommand{\Gm}{\mu}
\newcommand{\Gn}{\nu}

\newcommand{\Gs}{\sigma}

\newcommand{\Gth}{\theta}

\newcommand{\cA}{{\cal A}}
\newcommand{\cF}{{\cal F}}

\newcommand{\cL}{{\cal L}}

\newcommand{\cM}{{\cal M}}

\newcommand{\cV}{{\cal V}}

\newcommand{\ft}[2]{{\textstyle {\frac{#1}{#2}} }}

\newcommand{\dd}{\partial}

\newcommand{\g}{{\mathfrak{g}}}
\newcommand{\h}{{\mathfrak{h}}}

\newcommand{\k}{{\mathfrak{k}}}

\newcommand{\e}{{\mathfrak{e}}_{10}}
\newcommand{\Rn}{{\mathbb{R}}}

\newcommand{\E}{E_{10}}
\newcommand{\SL}{{\rm SL} (10, {\Rn})}

\newcommand{\be}{\begin{equation}}
\newcommand{\ee}{\end{equation}}
\newcommand{\ben}{\begin{displaymath}}
\newcommand{\een}{\end{displaymath}}
\newcommand{\ba}{\begin{eqnarray}}
\newcommand{\ea}{\end{eqnarray}}
\newcommand{\nn}{\nonumber}
\newcommand{\non}{\nonumber\\}

\newcommand{\la}{\label}

\newcommand{\Ref}[1]{(\ref{#1})}

\def\moth{\mathsurround=0pt}
\newdimen\zo \zo=0pt

\def\tick{\leaders\hrule height 0.5ex depth 0pt \hskip 0.5pt}
\def\upboxfill{$\moth \setbox\zo\hbox{\tick}%
  \hskip 2pt\hbox to 0pt{$\tick$\hss}\hrulefill \hbox to 6pt{$\tick$\hss}$}

\def\dtick{\leaders\hrule height .34pt depth .5ex \hskip 0.5pt}
\def\downboxfill{$\moth \setbox\zo\hbox{\dtick}%
  \hskip 2pt\hbox to 0pt{$\dtick$\hss}\hrulefill \hbox to 6pt{$\dtick$\hss}$}

\begin{document}
\draft
\preprint{AEI-2002-054, IHES/P/02/48}
\title{$\E$ and a ``small tension expansion'' of M Theory}
\author{T.~Damour}
\address{
Institut des Hautes Etudes Scientifiques, 35 route de Chartres,
F-91440 Bures-sur-Yvette, France
}
\author{M.~Henneaux}
\address{
Physique Th\'eorique et Math\'ematique, Universit\'e Libre de Bruxelles,
C.P.231, B-1050 Bruxelles, Belgium
}
\author{H.~Nicolai}
\address{
Max-Planck-Institut f\"ur Gravitationsphysik,
M\"uhlenberg 1, D--14476 Golm, Germany
}
\date{\today}
\maketitle
\begin{abstract}
A formal ``small tension'' expansion
of $D\!=\!11$ supergravity near a spacelike singularity
is shown to be equivalent, at least up to 30th order in height,
to a null geodesic motion in the infinite dimensional coset space
$\E/K(\E)$, where $K(\E)$ is the maximal compact subgroup of the
hyperbolic Kac-Moody group $\E (\Rn)$. For the proof we make use of
a novel decomposition of $\E$ into irreducible representations of its
$\SL$ subgroup. We explicitly show how to identify the first four 
rungs of the $\E$ coset fields with the values of geometric quantities
constructed from $D=11$ supergravity fields and their spatial
gradients taken at some comoving spatial point.
\end{abstract}

\begin{multicols}{2}
\narrowtext

The consideration of limits where some (possibly dimensionful)
parameter is taken to be small is often a way of revealing the
hidden symmetry structure of physical theories. In \cite{Gross},
it was argued that the small tension limit $T_s \! \rightarrow \! 0$ 
of string theory gives rise to an infinite number of relations
between string scattering amplitudes, indicating the presence of
an enormous symmetry. In this Letter, we shall consider the
bosonic sector of M Theory, and more specifically its low energy
limit, $D\!=\!11$ supergravity \cite{CJS}, in a limit which can
likewise be (intuitively) thought of as a small tension limit 
$T_b \! \rightarrow \! 0$, where $T_b := c^4 (32\pi G_N)^{-1}$ is 
the bulk tension governing the propagation of small excitations 
(e.g. gravitational waves) in the ten-dimensional spatial geometry.
Indeed, taking $T_b\!\rightarrow \! 0$ in the linearized
Einstein-Hilbert action $S=\frac12 \int dT\, d^{10}x \big( \rho_b
(\dd_T h_{ij})^2 - T_b (\dd_x h_{ij})^2\big)$ is equivalent to
taking the limit of vanishing velocity of propagation
$c=\sqrt{T_b/\rho_b}$; alternatively, it may be viewed as a strong
coupling limit ($G_N\! \rightarrow\!\infty$) \cite{IshamCT}. 
Physically, this limit is realized
near a spacelike singularity, where the different spatial points 
become causally disconnected as the horizon scale $\ell_H \sim cT$ 
becomes smaller than their spacelike separation ($T$ being the 
proper time), provided the time derivatives of the fields dominate 
their spatial gradients. As shown recently \cite{DH1,DH2},
this is indeed the case for the massless bosonic sector of 
$D\!=\!11$ supergravity. Furthermore, as $T_b \! \rightarrow \! 0$, 
the metric exhibits the chaotic oscillations originally discovered 
by Belinskii, Khalatnikov and Lifshitz (BKL) for the generic 
cosmological solution to Einstein's equations in four dimensions \cite{BKL}. 
The oscillatory evolution of the metric at each spatial point 
can be asymptotically described as a relativistic billiard taking 
place in the fundamental Weyl chamber of some indefinite 
Kac-Moody (KM) algebra \cite{DH1,DH2}. Chaos occurs when this KM algebra 
is hyperbolic, in particular for $\E$ \cite{DHJN}.

In this Letter we extend these tantalizing results much beyond the 
leading order by relating a BKL-type expansion to an algebraic 
expansion in the height of the positive roots of the Lie algebra 
of $\E$. We show how to map, up to height 30, geometrical objects 
of M theory onto coordinates in the infinite-dimensional coset space 
$\E/K(\E)$, where $K(\E$) is the maximal compact subgroup of the 
canonical real form of $\E$. Under this correspondence, the 
time evolution of the geometric M Theory data at each spatial 
point is mapped, up to height 30, onto some (constrained) null 
geodesic motion of $\E/K(\E)$. Our results underline the potential 
importance of $\E$, whose appearance in the reduction of $D\!=\!11$ 
supergravity to one dimension had been conjectured already long ago 
by B.~Julia \cite{Julia}, as a candidate symmetry underlying 
M theory (see also \cite{Mizoguchi}, and \cite{West} where 
$E_{11}$ was proposed as a fundamental symmetry of M Theory).

Introducing a zero-shift slicing ($N^i=0$) of the eleven-dimensional
spacetime, and a {\em time-independent} spatial zehnbein
$\Gth^a(x) \equiv {E^a}_i(x) dx^i$, the metric and four form
$\cF = d\cA$ become
\ba\label{Gauge}
  &&  ds^2 = - N^2 (d{x^0})^2 + G_{ab} \Gth^a \Gth^b \\ \!\!\!\!\!\!
\cF &=& \frac1{3!}\cF_{0abc}\,  dx^0 \!\wedge\!\Gth^a\!\wedge\!
\Gth^b\!\wedge\! \Gth^c
+ \frac1{4!}\cF_{abcd} \, \Gth^a\!\wedge\!\Gth^b\!\wedge\!
\Gth^c\!\wedge\!\Gth^d \nn
\ea
We choose the time coordinate $x^0$ so that the lapse $N=\sqrt{G}$,
with $G:= \det G_{ab}$ (note that $x^0$ is not the proper time
$T = \int N dx^0$;
rather, $x^0\rightarrow\infty$ as $T \rightarrow 0$). In this frame
the complete evolution equations of $D=11$ supergravity read
\ba\label{EOM}
\dd_0 \big( G^{ac} \dd_0 G_{cb} \big)  &=&
\ft16 G \cF^{a\Gb\Gg\Gd} \cF_{b\Gb\Gg\Gd} -
\ft1{72} G \cF^{\Ga\Gb\Gg\Gd} \cF_{\Ga\Gb\Gg\Gd} \Gd^a_b \non
   &&        - 2 G {R^a}_b (\GG,C) \non
\dd_0 \big( G\cF^{0abc}\big) &=&
\ft1{144} \varepsilon^{abc a_1 a_2 a_3 b_1 b_2 b_3 b_4}
          \cF_{0a_1 a_2 a_3} \cF_{b_1 b_2 b_3 b_4} \non
  &&  \!\! \!\! \!\! \!\! \!\! \!\! \!\! \!\!
 + \ft32 G \cF^{de[ab} {C^{c]}}_{de} - G {C^e}_{de} \cF^{dabc}
     - \dd_d \big( G\cF^{dabc} \big) \non
\dd_0 \cF_{abcd} &=& 6 \cF_{0e[ab} {C^e}_{cd]} + 4 \dd_{[a} \cF_{0bcd]}
\ea
where $a,b \in \{1,\dots,10\}$ and $\Ga,\Gb \in \{0,1,\dots,10\}$,
and $R_{ab}(\GG,C)$ denotes the spatial Ricci tensor; the (frame)
connection components are given by
$
2 G_{ad} {\GG^d}_{bc} = C_{abc} + C_{bca} - C_{cab} +
          \dd_b G_{ca} + \dd_c G_{ab} - \dd_a G_{bc}
$
with ${C^a}_{bc} \equiv G^{ad} C_{dbc}$ being the structure coefficients of
the zehnbein $d\Gth^a = \frac12 {C^a}_{bc} \Gth^b \!\wedge\! \Gth^c$.
The frame derivative is $\dd_a \equiv {E^i}_a (x) \dd_i$ (with
$ {E^a}_i {E^i}_b = \Gd^a_b$). To determine the solution at any {\it given}
spatial point $x$ requires knowledge of an infinite tower
of spatial gradients: one should thus
augment \Ref{EOM} by evolution equations for
$\dd_a G_{bc}, \dd_a \cF_{0bcd}, \dd_a \cF_{bcde}$, etc.,
which in turn would involve higher and higher spatial gradients.

The geodesic Lagrangian on $\E/K(\E)$ is defined by generalizing
the standard Lagrangian on a finite dimensional coset space $G/K$,
where $K$ is a maximal compact subgroup of the Lie group $G$.
All the elements entering the construction of $\cL$ have natural
generalizations to the case where $G$ is the group obtained by
exponentiation of a hyperbolic KM algebra. We refer readers
to \cite{Kac} for basic definitions and results of the theory of
KM algebras, and here only recall that a KM algebra $\g \equiv \g (A)$
is generally defined by means of a Cartan matrix $A$ and a set
of generators $\{e_i,f_i,h_i\}$ and relations (Chevalley-Serre
presentation), where $i,j=1,\dots, r\equiv {\rm rank}\, \g (A)$.
The elements $\{ h_i \}$ span the Cartan subalgebra (CSA)
$\h$, while the $e_i$ and $f_i$
generate an infinite tower of raising and lowering operators, respectively.
The ``maximal compact'' subalgebra $\k$ is defined as the subalgebra
of $\g (A)$ left invariant under the Chevalley involution
$\omega (h_i) = -h_i\, , \, \omega (e_i)= -f_i\,,\, \omega (f_i)= -e_i$.
In other words, $\k$ is spanned by the ``antisymmetric'' elements
$E_{\Ga,s} - E_{\Ga,s}^T $, where
$E_{\Ga,s}^T \equiv - \omega (E_{\Ga,s})$ is the ``transpose'' of
some multiple
commutator $E_{\Ga,s}$ of the $e_i$'s associated with the root $\Ga$ (i.e.
$[ h, E_{\Ga,s}] = \Ga (h)  E_{\Ga,s}$ for $h\in \h$). Here
$s=1,\dots {\rm mult} (\Ga)$ labels the different elements of $\g (A)$
having the same root $\Ga$.

The $\Gs$-model is formulated in terms of a one-parameter dependent group
element $\cV =\cV(t) \in \E$ and its Lie algebra valued derivative
\be
v(t) := \frac{d\cV}{dt} \cV^{-1} (t) \in \e \equiv {\rm Lie} \, \E
\ee
In physical terms, $\cV$ can be thought of as a vast extension of the
vielbein of general relativity (an ``$\infty$-bein''), and $\E$
and $K(\E)$ as infinite dimensional generalizations of the $GL(d,\Rn)$
and local Lorentz symmetries of general relativity. The action is
$\int dt \cL$ with
\be\label{Lag0}
\cL := {n(t)}^{-1} \langle v_{\rm sym}(t) | v_{\rm sym} (t)\rangle
\ee
with a ``lapse'' function $n(t)$ (not to be confused with $N$),
whose variation gives rise to the Hamiltonian constraint ensuring
that the trajectory is a null geodesic.
The ``symmetric'' projection $v_{\rm sym}:= \ft12 (v + v^T)$ eliminates
the component of $v$ corresponding to a displacement ``along $\k$'',
thereby defining an evolution on the coset space $\E/K(\E)$.
$\langle.|.\rangle$ is the standard invariant bilinear form on
the KM algebra \cite{Kac}. We note the existence of transcendental
KM invariants \cite{KP} that might be added to \Ref{Lag0} to represent
non-perturbative effects.

Because no closed form construction exists for the raising operators
$E_{\Ga,s}$, nor their invariant scalar products
$\langle E_{\Ga,s} | E_{\Gb,t} \rangle = N^\Ga_{s,t} \Gd^0_{\Ga + \Gb}$,
we have devised a recursive approach based on the decomposition
of $\E$ into irreducible representations of its $\SL$ subgroup.
Let $\Ga_1,\dots,\Ga_9$ be the nine simple roots of $A_9 \equiv sl(10)$
corresponding to the horizontal line in the $\E$ Dynkin diagram,
and $\Ga_0$ the ``exceptional'' root connected to $\Ga_3$. Its dual
CSA element $h_0$ enlarges $A_9$ to the Lie algebra of ${\rm GL} (10)$.
Any positive root of $\E$ can be written as
\be\label{E10root}
\Ga = \ell \Ga_0 + \sum_{j=1}^9 m^j \Ga_j \quad (\ell,m^j \geq 0)
\ee
We call $\ell \equiv \ell (\Ga)$ the ``level'' of the root $\Ga$.
This definition differs from the usual one, where the (affine) level 
is identified with $m^9$ and thus counts the number of appearances 
of the over-extended root $\Ga_9$ in $\Ga$\cite{FF,KMW}. Hence, 
our decomposition corresponds to a slicing (or ``grading'') of the
forward lightcone in the root lattice by spacelike hyperplanes, with only
finitely many roots in each slice, as opposed to the lightlike slicing
for the $E_9$ representations (involving not only infinitely many roots
but also infinitely many affine representations for $m^9\geq 2$ \cite{FF,KMW}).

The adjoint action of the $A_9$ subalgebra leaves the level $\ell(\Ga)$
invariant. The set of generators corresponding to a given level $\ell$
can therefore be decomposed into a (finite) number of irreducible
representations of $A_9$. The multiplicity of $\Ga$ as a root of
$\E$ is thus equal to the sum of its multiplicities as a weight
occurring in the $\SL$ representations. Each irreducible representation
of $A_9$ can be characterized by its highest weight $\GL$, or
equivalently by its Dynkin labels $(p_1,\dots,p_9)$ where
$p_k (\GL) := (\Ga_k,\GL)\geq 0$ is the number of columns
with $k$ boxes in the associated Young tableau. For instance, the
Dynkin labels $(001 000 000)$ correspond to a Young tableau consisting
of one column with three boxes, {\it i.e.} the antisymmetric tensor
with three indices. The Dynkin labels are related to the 9-tuple of
integers $(m^1,\dots,m^9)$ appearing in \Ref{E10root} (for the
highest weight $\GL\equiv - \Ga$) by
\be\la{mi}
 S^{i3} \ell - \sum_{j=1}^9 S^{ij} p_j = m^i \geq 0
\ee
where $S^{ij}$ is the inverse Cartan matrix of $A_9$. This relation
strongly constrains the representations that can appear at level $\ell$,
because the entries of $S^{ij}$ are all positive, and the 9-tuples
$(p_1,\dots,p_9)$ and $(m_1,\dots, m_9)$ must both consist of
non-negative integers. In addition to satisfying the Diophantine
equations \Ref{mi}, the highest weights must be roots of $\E$, which
implies the inequality
\be\la{L2}
\GL^2 = \Ga^2 =
 \sum_{i,j=1}^9 p_i S^{ij} p_j - \ft1{10} \ell^2 \leq 2
\ee
All representations occurring at level $\ell +1$ are contained in
the product of the level-$\ell$ representations with the $\ell=1$
representation. Imposing the Diophantine inequalities \Ref{mi}, \Ref{L2} 
allows one to discard many representations appearing in this product.
The problem of finding a completely explicit and
manageable representation of $\E$ in terms of an infinite tower of
$A_9$ representations is thereby reduced to the problem of determining
the outer multiplicities of the surviving  $A_9$ representations,
namely the number of times each
representation appears at a given level $\ell$. The Dynkin labels
(all appearing with outer multiplicity one)
for the first six levels of $\E$ are
\ba\la{irreps}
\ell=1  \quad &:& \quad (001000000) \non
\ell=2  \quad &:& \quad (000001000) \non
\ell=3  \quad &:& \quad (100000010) \non
\ell=4  \quad &:& \quad (001000001) \; , \; (200000000) \non
\ell=5  \quad &:& \quad (000001001) \; , \; (100100000) \non
\ell=6  \quad &:& \quad (100000011) \; , \; (010001000) \, ,\non
              &&            \quad (100000100) \; , \; (000000010)
\ea
The level $\ell \leq 4$ representations can be easily determined by
comparison with the decomposition of $E_8$ under its $A_7$ subalgebra
(see \cite{CJLP,KNS}) and use of the Jacobi identity, which
eliminates the representations $(000000001)$ at level three and
$(010000000)$ at level four. By use of a computer and the $\E$
root multiplicities listed in \cite{KMW,BGN}, the
calculation can be carried much further \cite{F}.

{}From \Ref{irreps} we can now directly read off the ${\rm} GL(10)$ tensors
making up the low level elements of $\E$. At level zero, we have the
$GL(10)$ generators ${K^a}_b$ obeying
$[{K^a}_b,{K^c}_d] = {K^a}_d \Gd^c_b -  {K^c}_b \Gd^a_d$. The $\e$
elements at levels $\ell=1,2,3$ are the $GL(10)$ tensors $E^{a_1a_2 a_3}$,
$E^{a_1 \dots a_6}$ and $E^{a_0|a_1 \dots a_8}$ with the symmetries
implied by the Dynkin labels (for the first three levels these
representions occur for all $E_n$, see \cite{OPR,West}).
The $\Gs$-model associates to these generators a corresponding tower
of ``fields'' (depending only on the ``time'' $t$): a zehnbein ${h^a}_b (t)$
at level zero, a three form $A_{abc} (t)$ at level one, a six-form
$A_{a_1\dots a_6} (t)$ at level two, a Young-tensor $A_{a_0|a_1 \dots a_8}(t)$
at level 3, etc. Writing the
generic $\E$ group element in Borel (triangular) gauge as
$ \cV(t) = \exp X_h (t) \cdot \exp X_A (t)$
with $X_h(t) = {h^a}_b  {K^b}_a$ and $X_A (t) = \ft1{3!} A_{abc} E^{abc}
+ \ft1{6!} A_{a_1\dots a_6} E^{a_1\dots a_6} +
\ft1{9!} A_{a_0|a_1 \dots a_8} E^{a_0|a_1 \dots a_8} + \dots$, and using the
$\E$ commutation relations in $GL(10)$ form together with the bilinear
form for $\E$, we find up to third order in level
\ba\label{Lag}
n \cL &=& \ft14 (g^{ac} g^{bd} - g^{ab} g^{cd}) \dot g_{ab} \dot g_{cd}
  + \ft12 \ft1{3!} DA_{a_1a_2a_3}DA^{a_1a_2a_3} \non
&&  \!\!\!\!\!\!\!\! \!\!\!
 + \ft12 \ft1{6!} DA_{a_1 \dots a_6}DA^{a_1\dots a_6}
  + \ft12 \ft1{9!} DA_{a_0 | a_1 \dots a_8} DA^{a_0|a_1\dots a_8}
\ea
where $g^{ab} = {e^a}_c {e^b}_c$ with $ {e^a}_b \equiv {(\exp h)^a}_b$,
and all ``contravariant indices'' have been raised by $g^{ab}$. The
``covariant'' time derivatives are defined by (with $\dd A\equiv \dot A$)
\ba\label{Dtime}
DA_{a_1a_2a_3} &:=& \dd A_{a_1a_2 a_3} \\
DA_{a_1\dots a_6} &:=& \dd A_{a_1 \dots a_6}
    + 10 A_{[a_1a_2 a_3} \dd A_{a_4a_5 a_6]} \non
DA_{a_1|a_2\dots a_9} &:=& \dd A_{a_1|a_2 \dots a_9}
    + 42 A_{\langle a_1a_2 a_3} \dd A_{a_4 \dots  a_9 \rangle} \non
&&  \!\!\!\! \!\!\!\!\!\!\!\! \!\!\!\!\!\!\!\! \!\!\!\!\!\!\!\! \!\!\!\!\!\!
- 42 \dd A_{\langle a_1a_2 a_3} A_{a_4 \dots  a_9 \rangle}
    + 280 A_{\langle a_1a_2 a_3} A_{a_4a_5 a_6} \dd A_{a_7a_8 a_9\rangle}
\nonumber
\ea
Here antisymmetrization $[\dots]$, and projection on the $\ell = 3$
representation $\langle \dots \rangle$, are  normalized with strength one
(e.g. $[[\dots]] = [\dots]$). Modulo field redefinitions,
all numerical coefficients in \Ref{Lag} and \Ref{Dtime} are uniquely
fixed by the structure of $\E$. Our expressions are reminiscent of
similar algebraic constructions in \cite{CJLP} and \cite{West}.
However, this is the first time that an algorithmic scheme based
on a Lagrangian in terms of the
invariant bilinear form on the hyperbolic KM algebra has been
proposed and worked out to low orders. Likewise, the general formulas
\Ref{mi} and \Ref{L2}, and the higher level representations in
\Ref{irreps} have not been exhibited before.

The Lagrangian \Ref{Lag0} is invariant under a nonlinear
realization of $\E$ such that $\cV(t) \rightarrow k_g(t) \cV(t) g$ with
$g\in\E$; the compensating ``rotation'' $k_g(t)$  being, in general,
required to restore
the ``triangular gauge''. When $g$ belongs to the nilpotent
subgroup generated by the $E^{abc}$, etc., this symmetry reduces to
the rather obvious ``shift'' symmetries of \Ref{Lag}
and no compensating rotation is needed. The latter are, 
however, required for the transformations generated by 
$F_{abc} = (E^{abc})^T$, {\it etc}. The associated infinite
number of conserved (Noether) charges are formally 
given by $J=\cM^{-1} \dd \cM$, where $\cM \equiv {\cV}^T \cV$. 
This can be formally solved in closed form as
\be\label{M}
\cM (t) = \cM (0) \cdot \exp (tJ)
\ee
The compatibility between \Ref{M} (indicative of the integrability 
of \Ref{Lag}) and the chaotic behavior of $g_{ab}(t)$ near a 
spacelike singularity will be discussed elsewhere.

The main result that we report in this letter is the following:
there exists a map between geometrical quantities constructed
at a given spatial point $x$ from the supergravity fields
$G_{\mu\nu}(x^0,x)$ and $\cA_{\mu\nu\rho}(x^0,x)$ and the
one-parameter-dependent quantities $g_{ab}(t), A_{abc} (t), \dots$ 
entering the coset Lagrangian \Ref{Lag}, under which the supergravity 
equations of motion \Ref{EOM} become equivalent, up to 30th order in 
height, to the Euler-Lagrange equations of \Ref{Lag}.
In the gauge \Ref{Gauge} this map is defined by
$t = x^0  \equiv \int dT/ \sqrt{G}$ and
\ba\label{map}
g_{ab}(t) &=& G_{ab} (t,x) \\
DA_{a_1a_2a_3}(t)  &=& \cF_{0a_1 a_2 a_3} (t,x) \non
DA^{a_1 \dots a_6} (t) &=&   - \ft1{4!}
\varepsilon^{a_1\dots a_6 b_1 b_2 b_3 b_4} \cF_{b_1 b_2 b_3 b_4} (t,x) \non
DA^{b|a_1 \dots a_8 } (t)&=& \ft32 \varepsilon^{a_1\dots a_8 b_1 b_2}
\big( {C^b}_{b_1 b_2} (x) + \ft29 \Gd^b_{[b_1}  {C^c}_{b_2] c} (x) \big)
\nonumber
\ea
The expansion in height ${\rm ht} (\Ga) \equiv \ell + \sum m^j$, which
controls the iterative validity of this equivalence, is as follows:
the Hamiltonian constraint of the coset model \Ref{Lag} contains an
infinite series of exponential coefficients $\exp\big(-2\Ga(\Gb)\big)$,
where $\Ga$ runs over all positive roots of $\E$, and where
$\Gb^a\equiv - {h^a}_a$ parametrize the CSA of $\E$. Previous work
has shown that, near a spacelike singularity ($t \rightarrow \infty$),
the dynamics of the
supergravity fields and of truncated versions of the $\E$ coset
fields is asymptotically dominated by the (hyperbolic) Toda model
defined by keeping only the exponentials involving the simple
roots of $\E$. Higher roots introduce smaller and smaller
corrections as $t$ increases. The ``small tension expansion'' of 
the equations of motion is then technically defined as a formal
BKL-like expansion that corresponds to such an expansion in
decreasing exponentials of the Hamiltonian constraint. On the
supergravity side, this expansion amounts to an expansion in 
gradients of the fields in appropriate frames. Level one 
corresponds to the simplest one-dimensional reduction of
\Ref{EOM}, obtained by assuming that both $G_{\Gm \Gn}$ and
$\cA_{\Gl \Gm \Gn}$ depend only on time \cite{DH1}; levels 2 and
3 correspond to configurations of $G_{\Gm \Gn}$ and $\cA_{\Gl \Gm \Gn}$ 
with a more general, but still very restricted $x$-dependence, 
so that {\it e.g.} the frame derivatives of the electromagnetic field 
in (\ref{EOM}) drop out \cite{DHHS}. When neglecting terms 
corresponding to ${\rm ht} (\Ga)\geq 30$, the map \Ref{map} 
provides a {\it perfect match} between the supergravity evolution 
equations \Ref{EOM} and the $\E$ coset ones, as well as
between the associated Hamiltonian constraints. (In fact, the 
matching extends to {\it all real roots} of level $\leq 3$.)

It is natural to view our map as embedded in a hierarchical 
sequence of maps involving more and more spatial gradients 
of the basic supergravity fields. Our BKL-like expansion 
would then be a way of revealing step by step a hidden 
hyperbolic symmetry, implying the existence of a 
huge non-local symmetry of Einstein's theory and its
generalizations. Although the validity of this conjecture remains
to be established, we can at least show that there is ``enough
room'' in $\E$ for all the spatial gradients. Namely, the search for
affine roots (with $m^9 =0$) in \Ref{mi} and \Ref{L2} reveals 
three infinite sets of admissible $A_9$ Dynkin labels
$(00100000n), (00000100n)$ and $(10000001n)$ with highest
weights obeying $\GL^2 =2$, at levels $\ell=3n+1,3n+2$ and $3n+3$, 
respectively. These correspond to three infinite towers of $\e$ elements 
\be\label{affine} 
{E_{a_1\dots a_n}}^{b_1 b_2b_3} \; , \;
{E_{a_1\dots a_n}}^{b_1 \dots b_6} \; , \; {E_{a_1\dots
a_n}}^{b_0| b_1 \dots b_8} 
\ee
which are symmetric in the lower indices and all appear with 
outer multiplicity one (together with three transposed
towers). Restricting the indices to $a_i=1$ and $b_i \in
\{2,...,10 \}$  and using the decomposition ${\bf 248}
\!\rightarrow\! {\bf 80}\!+ \!{\bf 84}\! +\! \overline{\bf 84}$ of
$E_8$ under its ${\rm SL}(9)$ subgroup one easily recovers the
affine subalgebra $E_9\subset\E$. The appearance of higher order
dual potentials ({\it \`a la} Geroch) in the $E_9$-based linear
system for $D\!=\!2$ supergravity \cite{BM} indeed suggests that
we associate the $\E$ Lie algebra elements \Ref{affine} to the
higher order spatial gradients $\dd^{a_1} \cdots \dd^{a_n} A_{b_1
b_2 b_3}, \dd^{a_1} \cdots \dd^{a_n} A_{b_1 \dots b_6}$ and
$\dd^{a_1} \cdots \dd^{a_n} A_{b_0|b_1 \dots b_8}$ or to some of
their non-local equivalents. Of course, the elements \Ref{affine}
generate only a tiny subspace of $\e$, suggesting the existence of
further M theoretic degrees of freedom and corrections beyond $D\!=\!11$
supergravity. Finally, we note that our approach based on a height
expansion can be extended to other physically relevant KM
algebras, such as $BE_{10}$ \cite{DH2,CJLP2} and $AE_n$
\cite{DHJN}.

We thank A.~Feingold, T.~Fisch\-bacher and V.~Kac for 
informative discussions.

\end{multicols}
\end{document}